\documentclass[reprint,aps,prl,superscriptaddress,footinbib,longbibliography]{revtex4-1}
\usepackage{amsmath,amssymb}
\usepackage{graphicx}
\usepackage{dcolumn}
\newcolumntype{d}[1]{D{.}{.}{#1}}

\usepackage{bm}
\usepackage{epstopdf}
\usepackage{graphicx}
\usepackage{stmaryrd}
\usepackage{tikz}
\usepackage{lipsum}
\usepackage{pgf}
\usepackage{float}
\usepackage[colorlinks, linkcolor= blue, citecolor = blue, urlcolor=blue]{hyperref}
\usepackage{tabularx}
\usepackage[none]{hyphenat}
\usepackage{subfigure}

\definecolor{lightblue}{RGB}{0,170,255}

\newcommand{\appropto}{\mathrel{\vcenter{
			\offinterlineskip\halign{\hfil$##$\cr
				\propto\cr\noalign{\kern2pt}\sim\cr\noalign{\kern-2pt}}}}}

\usepackage[normalem]{ulem}
\newcommand{\dps}{\displaystyle}

\newcommand{\om}{\iffalse}
\newcommand{\pd}[2]{\frac{\partial #1}{\partial #2}}

\newcommand{\ba}{\arraycolsep 0.3ex \begin{array}{rl}}
	\newcommand{\ea}{\end{array}}
\newcommand{\bc}{\begin{cases}}
	\newcommand{\ec}{\end{cases}}
\usepackage{color}
\newcolumntype{C}[1]{>{\centering\arraybackslash}p{#1}}

\begin{document}
	\title{Quantum geometry induced second harmonic generation}
\author{Pankaj Bhalla}
	\email{bhalla@csrc.ac.cn}
	\affiliation{Beijing Computational Science Research Center, Beijing 100193, China}
	\affiliation{ARC Centre of Excellence in Future Low-Energy Electronics Technologies, The University of New South Wales, Sydney 2052, Australia}
	\author{Kamal Das}
	\email{kamaldas@iitk.ac.in}
	\affiliation{Department of Physics, Indian Institute of Technology Kanpur, Kanpur-208016, India}
\author{Dimitrie Culcer}
\email{d.culcer@unsw.edu.au}
	\affiliation{ARC Centre of Excellence in Future Low-Energy Electronics Technologies, The University of New South Wales, Sydney 2052, Australia}
	\affiliation{School of Physics, The University of New South Wales, Sydney 2052, Australia}
\author{Amit Agarwal}
	\email{amitag@iitk.ac.in}
	\affiliation{Department of Physics, Indian Institute of Technology Kanpur, Kanpur-208016, India}

\date{\today}

\begin{abstract} 
Quantum geometry of the electron wave-function plays a significant role in the linear and non-linear responses of crystalline materials. Here, we study quantum geometry induced second harmonic generation. We identify non-linear responses stemming from the quantum geometric tensor and the quantum geometric connection in systems with finite Fermi surfaces and disorder. In addition to the injection, shift, and anomalous currents we find two new contributions, which we term double resonant and higher-order pole contributions. Our findings can be tested in state-of-the-art devices in WTe$_2$ (time reversal symmetric system) and in CuMnAs (parity-time reversal symmetric systems).
\end{abstract}

\maketitle

The quantum geometric properties of the electron wave-function, such as the Berry curvature and the orbital magnetic moment, give rise to a multitude of electronic transport effects~\cite{Gao_Nature2021,Lai_NatNano2021,sundaram_PRB1999_wave,xiao_RMP2010_berry,nagaosa_RMP2010_anomalous,
son_PRB2012_berry, son_PRB2013_chiral,das_PRR2020_thermal,gao_PRL2014_field,gao_FP2019_semi,
papaj_PRL2019, mandal_PRB2020}. Likewise, in optical phenomena, the Berry phase plays a key role in several photogalvanic responses~\cite{Moore_2010PRL, sodemann_PRL2015}, in the circular quantised photogalvanic effect~\cite{juan_NC2017}, and in non-linear optical responses in systems with broken time reversal symmetry~\cite{holder_PRR2020, ahn_PRX2020, bhalla_PRL2020,  watanabe_PRX2021, watanabe_PRB2021}. Other geometric quantities such as the Berry curvature dipole and the quantum metric also play a fundamental role in determining non-linear optical effects \cite{khurgin_APL1995, Aversa_PRB1995, Sipe_PRB2000, glazov_PR2014, morimoto_SA2016, morimoto_PRB2016, khurgin_JOSAB2016, nagaosa_AM2017, nagaosa_NC2018, sirica_PRL2019, parker_PRB2019, bhalla_PRL2020, juan_PRX2020,Ahn_arXiv2021}. However, a concrete understanding of the connection between different geometric quantities and second harmonic (SH) generation~\cite{Ghimire_Scienc2011, Sipe_PRB1993, Sipe_PRB2000, passos_2018PRB, Yang_arxiv2018, li_PRB2018, takasan_arxiv2020}, higher harmonic generation~\cite{Lewenstein_PRA1994,Lysne_PRBL2021,kim_arxiv2021,juan_PRR2020}, and the photogalvanic effect~\cite{hosur_PRB2011, mikhailov_PRB2011,  zhang_PRB2018, rostami_NLAH_PRB18, you_PRB2018, golub_PRB2018, ma_NM2019, okamura_NC2020, Yoshikawa_Ncom2019,tan_PRB2019, inarrea_PRB2019, juan_PRX2020, wang_NPJCM2020,  fei_PRB2020, gao_arxiv2020,  Zou_AM2020,kaplan_arxiv2021,gao_PRB2021,wei_arxiv2021,chaudhary_arxiv2021,shi_PRL2021} has thus far been lacking. 

Here, we revisit the theory of SH generation in insulating and metallic systems, 
classify the different contributions according to their quantum geometrical origin, and identify two previously unknown SH responses induced by quantum geometric properties. We re-formulate the theory of SH generation \cite{Aversa_PRB1995, Sipe_PRB2000} within the quantum kinetic framework and establish its connection to different quantum geometric properties of the electronic wave-function. We explicitly i) demonstrate the importance of the quantum geometric quantities on the different contributions to the SH current ii) calculate the Fermi surface contributions to the SH current and iii) include the effect of disorder. We show that the SH current is primarily determined by four geometrical quantities, the Berry curvature ($\Omega$), the quantum metric ($\cal G$), the metric connection ($\Gamma$) and the symplectic connection ($\tilde \Gamma$). The former two are the real and imaginary part of the quantum geometric tensor (${\cal Q} = {\cal G} - i\Omega/2$) while the latter two define the quantum geometric connection (${\cal C} = \Gamma-i {\tilde\Gamma}$), respectively \cite{ahn_PRX2020}. 

We find five distinct contributions to the SH current, which are related to the quantum geometric properties of the electron wave-function [see Table~\ref{table_1}]. Three of these are the injection, shift and anomalous SH contributions, which are similar in nature to the corresponding photogalvanic (DC) counterparts. The other two, which we term double resonant and higher-order pole contributions, were not explored previously. 
Our detailed calculations show that in time reversal $(\cal T)$ symmetric systems, such as monolayer WTe$_2$, the shift contribution depends only on $\tilde{\Gamma}$, while the other four contributions rely only on $\Omega$. In contrast, in parity-time reversal symmetric ($\mathcal{P T}$) systems, such as CuMnAs, the shift current is determined by $\Gamma$, while all other geometric contributions are determined by $\cal G$. We show that the SH injection current vanishes in ${\cal T}$ symmetric systems, while all the current components are finite in ${\cal P}{\cal T}$ symmetric systems.

\begin{table*}[t]
	\caption{Different terms leading to second harmonic ($2 \omega$) current in response to a harmonic electromagnetic field [see Eq.~\eqref{SHG_cur}].  
	The SH conductivity is given by the relation, $j_a^{(2)}(2 \omega) = \sum_{bc}\sigma_{abc}(2\omega) E_b(\omega)E_c(\omega)$ and the SH conductivities are defined to be symmetric under the exchange of the last two indices. This is achieved via the relation, $\sigma_{abc} = \sigma_{acb} = -1/(2\pi)^{d}(e^3/\hbar^2) \int d^d k [{\cal I}_{abc} + {\cal I}_{acb}]/2$. We define the quantum metric as $\mathcal{G}_{mp}^{cb} = \{ {\mathcal R}^c_{pm}, {\mathcal R}_{mp}^b \}/2$, the Berry curvature as $\Omega_{mp}^{cb} = i[{\mathcal R}^c_{pm}, {\mathcal R}_{mp}^b]$ \cite{watanabe_PRX2021}, and the metric connection ($\Gamma$) and symplectic connection ($\tilde\Gamma$) terms as $\mathcal{R}_{pm}^{a}  \mathcal{D}^{b}_{mp}\mathcal{R}^c_{mp}=\Gamma_{mp}^{abc} - i \tilde\Gamma_{mp}^{abc} $, where $\mathcal{D}^b_{mp} = \partial_b - i(\mathcal{R}_{mm}^b - \mathcal{R}_{pp}^b)$.
	Corresponding to each SH current component, there is also a photogalvanic (DC) counterpart, which can be obtained via the substitution $g^{2\omega}_{mp} \to g^{\omega = 0}_{mp}$ and $g_0^{2\omega} \to g_0^{\omega=0}$ (see Table S1 in the SM~\cite{Note1}).} 
	
	\centering  
	{
		\begin{tabular}{c c c c c c }
			\hline \hline
			\rule{0pt}{3ex}
			  Current & Integrand & Geometrical quantity & $~{\cal T}~$ & $~{\cal PT}~$ &  Physical Origin \\ [1ex]
			\hline \hline
			\rule{0pt}{3ex}
			Drude ~~~&~~~ ${\mathcal I}^{\rm D}_{abc} =g_{0}^{\omega}g_{0}^{2\omega} \sum_m \frac{1}{\hbar}\frac{\partial\varepsilon_m}{\partial k_a} \frac{\partial^2f_{m}^{(0)}}{\partial k_b \partial k_c}$ & None & 0 & $\neq 0$ & Non eq. distribution function\\ [2ex] 
			Injection & ${\mathcal I}^{\rm Inj}_{abc} = - g^{2\omega}_{0} \sum_{pm} g^{\omega}_{mp}  \frac{\partial\omega_{mp}}{\partial k_a} (\mathcal{G}_{mp}^{bc} - i \Omega_{mp}^{bc}/2)  F_{mp}$  
			&~${\Omega}$, ${\cal G}$ & $\Omega$ 
			& ${\cal G}$ & Velocity injection along current
			\\ [2ex]
			 Shift & ${\mathcal I}^{\rm Sh}_{abc} = \sum_{pm} \omega_{mp} g_{mp}^{2\omega}g_{mp}^{\omega} (\Gamma_{mp}^{abc} - i\tilde \Gamma_{mp}^{abc}  ) F_{mp} $ & $\tilde\Gamma$, $\Gamma$ & $\tilde\Gamma$ & $\Gamma$ & Shift of the wave-packet\\[2ex]
			 \rule{0pt}{3ex}
			 Anomalous & ${\mathcal I}^{\rm An}_{abc} = g_{0}^{\omega} \sum_{pm} \omega_{mp}  g_{mp}^{2\omega}   (\mathcal{G}_{mp}^{ab} - i \Omega_{mp}^{ab}/2) \frac{\partial F_{mp}}{\partial k_c}$ & ${\Omega}$, ${\cal G}$ & ${\Omega}$ & ${\cal G}$& Fermi Surface \\ [2ex] 
			 \rule{0pt}{3ex} 
			 DR & ${\mathcal I}^{\rm DR}_{abc} = \sum_{pm}\omega_{mp} g_{mp}^{\omega} g_{mp}^{2\omega} (\mathcal{G}_{mp}^{ac} - i \Omega_{mp}^{ac}/2) \frac{\partial F_{mp}}{\partial k_b}$ & ${\Omega}$, ${\cal G}$ & ${\Omega}$ & ${\cal G}$ & Fermi surface \\[2ex] 
			 \rule{0pt}{3ex} 
			 HOP & ${\mathcal I}^{\rm HOP}_{abc} = \sum_{pm}\omega_{mp} g_{mp}^{2\omega} (\mathcal{G}_{mp}^{ac} - i \Omega_{mp}^{ac}/2) \frac{\partial g_{mp}^{\omega}}{\partial k_b} F_{mp}$ & ${\Omega}$, ${\cal G}$ & ${\Omega}$ & ${\cal G}$ & Velocity injection along field\\ [2ex] 
						\hline \hline
	\end{tabular}}
	\label{table_1}
\end{table*}

The quantum kinetic equation for the density matrix $\rho({\bm k},t)$ in the crystal momentum representation is 
\begin{equation}
\frac{\partial \rho ({\bm k},t)}{\partial t} + \frac{i}{\hbar}\left[\mathcal{H}({\bm k},t), \rho ({\bm k},t)\right] + \frac{\rho ({\bm k},t)}{\tau} =0~.
\end{equation}
Here, $\mathcal{H}({\bm k},t) = \mathcal{H}_0({\bm k}) + e{\bm r}\cdot{\bm E}(t)$ is the full Hamiltonian of the clean system, with $\mathcal{H}_0$ being the unperturbed band Hamiltonian. The last term represents the light-matter interaction in the length gauge with `$-e$' being the electronic charge. The external time dependent homogeneous electric field is given by ${\bm E}(t) = {\bm E}_0 e^{-i\omega t}$, with ${\bm E}_0 = \{E_0^x, E_0^y, E_0^z\}$ being the electric field strength and $\omega$ is the frequency. To simplify the analytical calculations we approximate the scattering term by $\rho/\tau$. The relaxation time $\tau$ accounts generically for impurity and phonon scattering, as well as recombination, and is assumed to be constant across the Fermi surface.

The nonlinear dynamics of the charge carriers can be explored by expanding the density matrix perturbatively in powers of the electric field, $\rho({\bm k},t) = \rho^{(0)}({\bm k},t) + \rho^{(1)}({\bm k},t) + \rho^{(2)}({\bm k},t) + \cdots$, where $\rho^{(N)}({\bm k},t) \propto (E_0^b)^{N}$. Expressing the density matrix in terms of the eigenstates of the unperturbed Hamiltonian, ${\cal H}_0 |u_{{\bm k}}^{p} \rangle = \varepsilon_{p,{\bm k}} |u_{{\bm k}}^{p} \rangle$, the kinetic equation for the $N^{\text{th}}$ order term in the density matrix, $\rho^{(N)}({\bm k},t) \equiv \rho^{(N)}$, is given by 
\begin{equation}
\pd{\rho_{mp}^{(N)}}{t} + \frac{i}{\hbar} \left[\mathcal{H}_0,\rho^{(N)}\right]_{mp} +  \frac{\rho^{(N)}_{mp}}{\tau} = \frac{e{\bm E}(t)}{\hbar} \cdot \left[{D}_{\bm k} \rho^{(N-1)}\right]_{mp}.
\label{eqn:den2}
\end{equation}
Here, we define the covariant derivative as $[{D}_{\bm k}\rho]_{mp} = \partial_{\bm k} \rho_{mp} -i[\mathcal{R}_{\bm k}, \rho]_{mp}$, where $\mathcal{R}_{mp}({\bm k}) = i \langle u_{\bm k}^{m} \vert \partial_{\bm k}u_{{\bm k}}^{p} \rangle$ is the momentum space non-Abelian Berry connection.

Up to linear-order in the external field strength, the solution of Eq.~(\ref{eqn:den2}) yields $\rho_{mp}^{(1)} = \sum_c \tilde{\rho}_{mp}^{(1,c)} E^c_0 e^{-i \omega t}$, where 
\begin{equation}
\tilde{\rho}_{mp}^{(1,c)}  = \frac{e}{\hbar}  g_{mp}^\omega \left[ \partial_{b} \rho^{(0)}_{mp} \delta_{mp} + i { \mathcal R}^c_{mp} F_{mp}\right].
\label{eqn:den4}
\end{equation}
Here, we have defined $F_{mp} \equiv f_{m}^{(0)} - f_{p}^{(0)}$ to be the difference between the occupation in bands $p$ and $m$ in equilibrium. The occupation of the bands is given by  $f_m^{(0)} \equiv \rho_{mm}^{(0)} = [1 + e^{\beta (\varepsilon_{m,{\bm k}}-\mu)}]^{-1}$  the Fermi-Dirac distribution function, where $\beta = 1/(k_BT)$, $k_B$ is the Boltzmann constant, $T$ is the absolute temperature, and $\mu$ is the chemical potential. The function $g_{mp}^{\omega} \equiv [1/\tau -i(\omega - \omega_{mp})]^{-1}$ with $ \hbar \omega_{mp} = (\varepsilon_{m,{\bm k}}-\varepsilon_{p,{\bm k}})$ is related to the joint density of states broadened by disorder, and $g_0^\omega = [1/\tau - i\omega]^{-1}$. The details of the calculations are given in Sec.~S1 of the Supplemental material (SM) 
\footnote{\href{https://www.dropbox.com/s/rky1pabnm2i58vh/supplementary.pdf?dl=0}{The Supplemental material} discusses, i) quantum kinetic theory to compute the electric field induced linear and second-order corrections to the density matrix, ii) symmetry properties of different  geometric quantities involved in the SH generation, iii) symmetry analysis of the different contributions to the SH currents, and iv) model specific calculations for $\mathcal{T}$ symmetric, and $\mathcal{P T}$ symmetric systems.}. 
In Eq.~\eqref{eqn:den4}, the first term captures the intra-band contributions ($\rho^{(0)}_{mp} = 0$ for $m\neq p$) which is finite only in the presence of a finite Fermi surface (e.g. doped semimetals and metals). The second term in Eq.~\eqref{eqn:den4} captures inter-band transitions as $F_{mp} = 0$ for $m=p$. 

Using the first order solution of the density matrix, $\rho_{mp}^{(1)}$, the second-order correction can be calculated to be, $\rho_{mp}^{(2)} = \sum_{bc}\tilde{\rho}_{mp}^{(2,bc)}E_0^b E_0^c e^{-i 2 \omega t}$. Here, 
\begin{equation}
\tilde{\rho}_{mp}^{(2,bc)} = \frac{e^2 g_{mp}^{2\omega}}{\hbar^2}  \Big[\partial_{b}\tilde{\rho}_{mp}^{(1,c)} - i \sum_{n} \Big(\mathcal{R}_{b}^{mn} \tilde{\rho}_{np}^{(1,c)} - \mathcal{R}_{b}^{np} \tilde{\rho}_{mn}^{(1,c)} \Big)\Big], 
\label{eqn:den5}
\end{equation} 
with the second term involving a sum over all the bands. We highlight that even the intra-band or diagonal terms ($m=p$) in $\rho_{mp}^{(2)}$ have contributions arising from the inter-band or off-diagonal terms in $\rho_{mp}^{(1)}$ (see Sec.~S1 of the SM~\cite{Note1}).

The time dependent nonlinear current is calculated from the definition, ${\bm j}(t) = -e \text{Tr}[\hat{{\bm v}} \rho(t)]$. In the eigenbasis of ${\cal H}_0$, the velocity operator $\hat{ v}^a$ can be expressed as ${ v}^a_{pm}({\bm k}) = \hbar^{-1} (\delta_{pm} \partial_{ k_a}\varepsilon_{m, {\bm k}} + i  \mathcal{R}^a_{pm} \hbar\omega_{pm})$. It includes the intra-band term in the form of the band velocity, and the inter-band term dependent on the non-Abelian Berry connection~\cite{wilczek_PRL1984_appear, culcer_PRB2005_coherent, chang_JPCM2008_berry}. The current can also be expressed as a power series of the electric field strength, ${\bm j}(t) = \sum_N {\bm j}^{(N)}$ with ${j}^{(N)} \propto (E_0^b)^{N}$. The SH component of the current is given by the $2\omega$ component of $j^{(2)}_a(t)$.  

Explicitly calculating the SH current for a $d$ dimensional system, we obtain 
\begin{equation} \label{SHG_cur}
j^{(2)}_a(t) = - \frac{e^3}{\hbar^2}e^{-i2\omega t} \sum_{bc} E^b_0 E^c_0 \int_{\rm BZ} \frac{d^d k}{(2\pi)^d} \mathcal{I}_{abc}({\bm k},\omega)~,
\end{equation}
where, $\mathcal{I}_{abc} = \mathcal{I}_{abc}^{\rm D} + \mathcal{I}_{abc}^{\rm Inj} + \mathcal{I}_{abc}^{\rm Sh} +  \mathcal{I}_{abc}^{\rm An} + \mathcal{I}_{abc}^{\rm DR} + \mathcal{I}_{abc}^{\rm HOP}$, and ${\rm BZ}$ denotes the Brillouin zone.
Here, based on the corresponding DC counterparts \cite{watanabe_PRX2021}, we have denoted the different SH contributions to the integrand as follows: Drude ($\mathcal{I}_{abc}^{\rm D}$), injection ($\mathcal{I}_{abc}^{\rm Inj}$), shift ($\mathcal{I}_{abc}^{\rm Sh}$), and anomalous ($\mathcal{I}_{abc}^{\rm An}$). In addition to these, we find two more contributions, which we refer to as the {\it double resonant} (DR) ($\mathcal{I}_{abc}^{\rm DR}$) and {\it higher-order pole} ($\mathcal{I}_{abc}^{\rm HOP}$) contributions. 
Both of these, explained below, depend on the scattering time $\tau$, and diverge if $\tau$ is unbounded. Below, we explicitly show that the DR and HOP contributions can be as large as the other contributions. The functional forms of all the contributions in the SH (photogalvanic) current are tabulated in Table \ref{table_1} (Table S1 of the SM~\cite{Note1}).

The DR current is a Fermi surface phenomenon, which shows resonant features for $\omega= \mu$ and $\omega =2\mu$ in a particle-hole symmetric system. The double resonance stems from the product of the joint density of states ($g_{mp}^\omega g_{mp}^{2\omega}$), reflecting the interplay of one-photon and two-photon absorption processes. The $g_{mp}^\omega$ factor denotes the single photon process contributing in the first order correction to the density matrix and it gives a peak at $\omega = 2 \mu$, in systems with particle-hole symmetry. 
The factor of $g_{mp}^{2\omega}$ is associated with two-photon absorption where the photon frequencies are additive, and this leads to a peak at $2\hbar \omega = 2\mu$. The HOP current, on the other hand, is a Fermi sea phenomenon, its name originating from the second-order poles in $\partial_b g_{mp}^\omega$. In addition to the second-order pole, the $\partial_b g_{mp}^\omega$ term also gives rise to velocity injection along the direction of the applied field, making the HOP contribution similar in spirit to the injection current.   
The HOP contribution is $\propto \tau, \tau^2$ for ${\cal T}$-symmetric and ${\cal P}{\cal T}$-symmetric systems, respectively. Among the three Fermi sea contributions, the injection current depends on the velocity difference between the two bands along the direction of current. The shift current is determined by the shift vector determining the positional shift of the carriers in real space, and the HOP current depends on the velocity difference parallel to the electric field.
            
Apart from the Drude current, which depends only on the band velocity, all the other components of current in Table~\ref{table_1} depend on the quantum geometric properties of the electron wave-function. Except for the shift current, four of these originate from the geometric quantity referred to as the {\it quantum geometric tensor} $Q_{mp}^{bc} = \mathcal{R}_{pm}^{b}\mathcal{R}_{mp}^{c} = {\mathcal G}^{bc}_{mp} - (i/2) \Omega^{bc}_{mp}$ \cite{Provost_CMP1980, Ma_PRB2010, Zhang_PRB2019}. Here, the non-Abelian Berry curvature is anti-symmetric under the exchange of spatial coordinates, $\Omega^{bc}_{mp}=-\Omega^{cb}_{mp}$, while the quantum metric is symmetric, $\mathcal{G}_{mp}^{bc} = \mathcal{G}_{mp}^{cb} $. Note that in the injection current the Cartesian indices of $\Omega$ and ${\cal G}$ are determined by the electric field direction.  However, in the other SH components, one of the indices of $\Omega$ and $\cal G$ is determined by the direction of the current. 
The anomalous part of the SH current is related to the Berry curvature and quantum metric. It is easily checked that the photogalvanic counterpart ($2\omega \to  0$) of the anomalous current can be expressed as a function of the Berry curvature dipole \cite{sodemann_PRL2015}. For the shift current one can define a third rank tensor, the {\it quantum geometric connection} ${\mathcal C}_{mp}^{abc} =\mathcal{R}_{pm}^{a}  \mathcal{D}_{mp}^{b}\mathcal{R}^c_{mp}$ with $\mathcal{D}^b_{mp} = \partial_b - i(\mathcal{R}_{mm}^b - \mathcal{R}_{pp}^b)$, which is symmetric under the interchange of the last two spatial indices. The quantum geometric connection is decomposed into real and imaginary parts ${\mathcal C}_{mp}^{abc}= \Gamma_{mp}^{abc} - i \tilde \Gamma_{mp}^{abc}$, where $\Gamma$ ($\tilde \Gamma$) is the metric (symplectic) connection \cite{Ahn_arXiv2021}.

Using the symmetry relations for different geometric quantities (see Sec.~S2 of the SM \cite{Note1}), we can easily check that all the SH current components  vanish in ${\cal P}$ symmetric systems. Furthermore, we find that in ${\cal T}$ symmetric systems, the SH current integrand corresponding to the i) Drude current, ii) the symplectic connection  part of the shift current, and iii) the quantum metric  part of the remaining four components are odd functions of ${\bm k}$. Thus, these contributions vanish in ${\cal T}$ symmetric systems after performing the BZ integration. Effectively, the injection, anomalous, DR and HOP contributions in ${\cal T}$ symmetric systems arise from the Berry curvature alone. 
%
For systems having ${\mathcal P} {\mathcal T}$ symmetry, the Berry curvature and symplectic connection vanish in the whole BZ and hence do not contribute  to the current. However, the quantum metric and the metric connection are non-zero and contribute to the different currents. Additionally, in ${\mathcal P} {\mathcal T}$ symmetric systems $\varepsilon({\bm k}) \neq \varepsilon(-{\bm k})$, and thus the non-linear Drude current can be finite in these systems. 

\begin{figure}[t!]
\includegraphics[width=\linewidth]{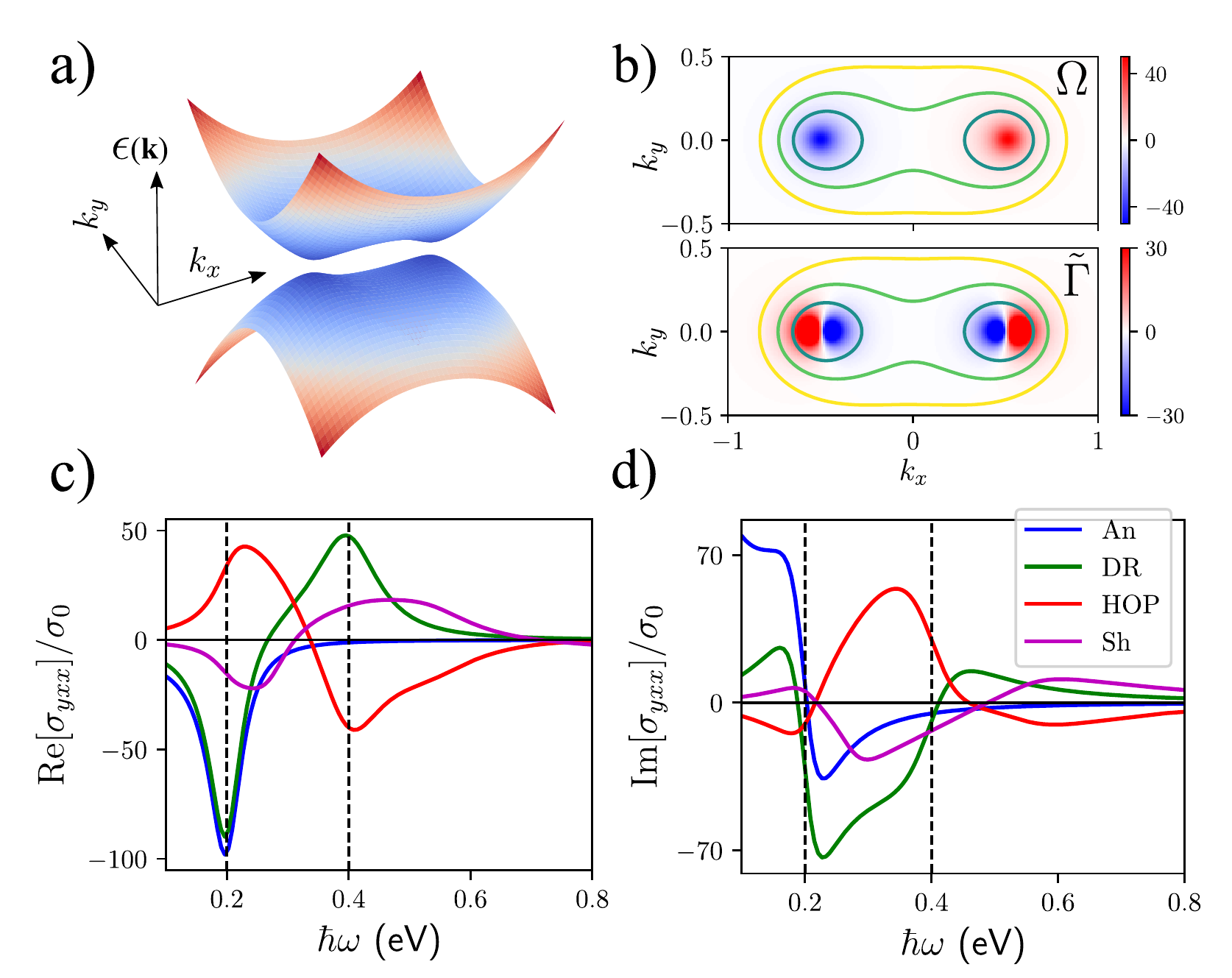}
\caption{a) Band dispersion for monolayer WTe$_2$ [from Eq.~\eqref{model_1}]. We have set the tilt $A=0$, gap parameters $\delta=-0.25, D=0.1$ and the other parameters to be $B=1.0,  v_y =1$ in eV. b) The momentum space distribution of $\Omega^{xy}_{cv}$ (top) and the symplectic  connection $\tilde\Gamma ^{yxx}_{cv}$ (bottom), where the subscript $cv$ denotes conduction and valence band. Frequency dependence of the different contributions to the c) real and d) imaginary parts of the conductivity $\sigma_{yxx}$. The conductivities are expressed in units of $\sigma_0= 10^{-5}$ ${\rm nA.m}/{\rm V}^2$. We have set $\mu=0.2$ eV, $\tau=1$ ps and temperature $T=12$ K.}
\label{fig_1}
\end{figure}
%

In $\mathcal{T}$ symmetric systems, the injection, anomalous, DR and HOP contributions depend only on the Berry curvature, whose diagonal components (in spatial indices) are zero. Thus, the corresponding diagonal elements of the conductivity tensor are zero ($\sigma^{\rm Inj/An/DR/HOP}_{aaa}=0$). For a 2D ${\cal T}$-symmetric system, there are only four independent components of $\sigma$: $\sigma_{yxx}, \sigma_{xyy}, \sigma_{xxy}$ and $\sigma_{yyx}$. Furthermore, since $\sigma^{\rm Inj}_{abc} \propto \Omega^{bc}$, we have $\sigma^{\rm Inj}_{abb} = 0$, and due to the anti-symmetric nature of the Berry curvature, we have $\sigma_{xxy}^{\rm Inj} =0$ and $\sigma_{yxy}^{\rm Inj} = 0$. Thus, in ${\cal T}$ symmetric  2D systems, the injection current does not  contribute to the SH generation. 
However, we expect non-zero SH contributions from the  anomalous, DR and HOP currents. Since $\sigma_{aaa}^{\rm {An, DR,HOP}}=0$, these components give rise to a Hall like response. For the shift current, which depends on the symplectic connection, we find $\sigma_{aaa}^{\rm Sh} \neq 0$ and this is the only finite contribution to the diagonal SH component. 

As an example of a ${\cal T}$-symmetric system, we consider a two band model of monolayer WTe$_2$ \cite{papaj_PRL2019, mandal_PRB2020}, 
\begin{equation}  \label{model_1}
\ba
\mathcal{H}({\bm k}) &\dps = A k^2 \openone + (B k^2 + \delta) \sigma_z + v_y k_y \sigma_y + \Delta\sigma_x~.
\ea
\end{equation}
Here, $A$ gives tilt to the dispersion, $v_y$ is the velocity component which gives rise to an anisotropic dispersion and $\Delta$ controls the band-gap. 
The $\cal H$ in Eq.~\eqref{model_1} has a mirror symmetry $M_x$, which enforces ${\mathcal H}(k_x, k_y) = {\mathcal H}(-k_x, k_y)$. 
The energy dispersion for this model is given by $\varepsilon({\bm k})=Ak^2 \pm \sqrt{(Bk^2 +\delta)^2 + k_y^2 v_y^2 +\Delta^2 }$. The corresponding band-structure is shown in Fig.~\ref{fig_1}(a)  with the corresponding geometric quantities, the Berry curvature and the metric connection, displayed in Fig.~\ref{fig_1} (b). Owing to the combination of mirror symmetry and ${\cal T}$-symmetry in Eq.~\eqref{model_1}, we have $\Omega_{mp}^{xy}(k_x,k_y) = \Omega_{mp}^{xy}(k_x,-k_y)$. This makes the BZ integrand for the SH components, ${\cal I}^{\rm An/DR/HOP}_{xyy} $ an odd function of $k_y$, and consequently $\sigma_{xyy}^{\rm An/DR/HOP} = 0$. However, the SH conductivity components $\sigma_{yxx}^{\rm An/DR/HOP}$ is finite and these generate a Hall current $j_y^{(2)} = \sigma_{yxx} E_x^2$. The real and imaginary parts of the different terms in the SH conductivity $\sigma_{yxx}$ are shown in Figs.~\ref{fig_1}(c), (d). 
The double resonant peak of $\sigma^{\rm DR}_{yxx}$ can be clearly seen in Fig.~\ref{fig_1}(c), along with the finite $\sigma^{\rm HOP}_{yxx}$ contribution.

\begin{figure}[t!]
\includegraphics[width=\linewidth]{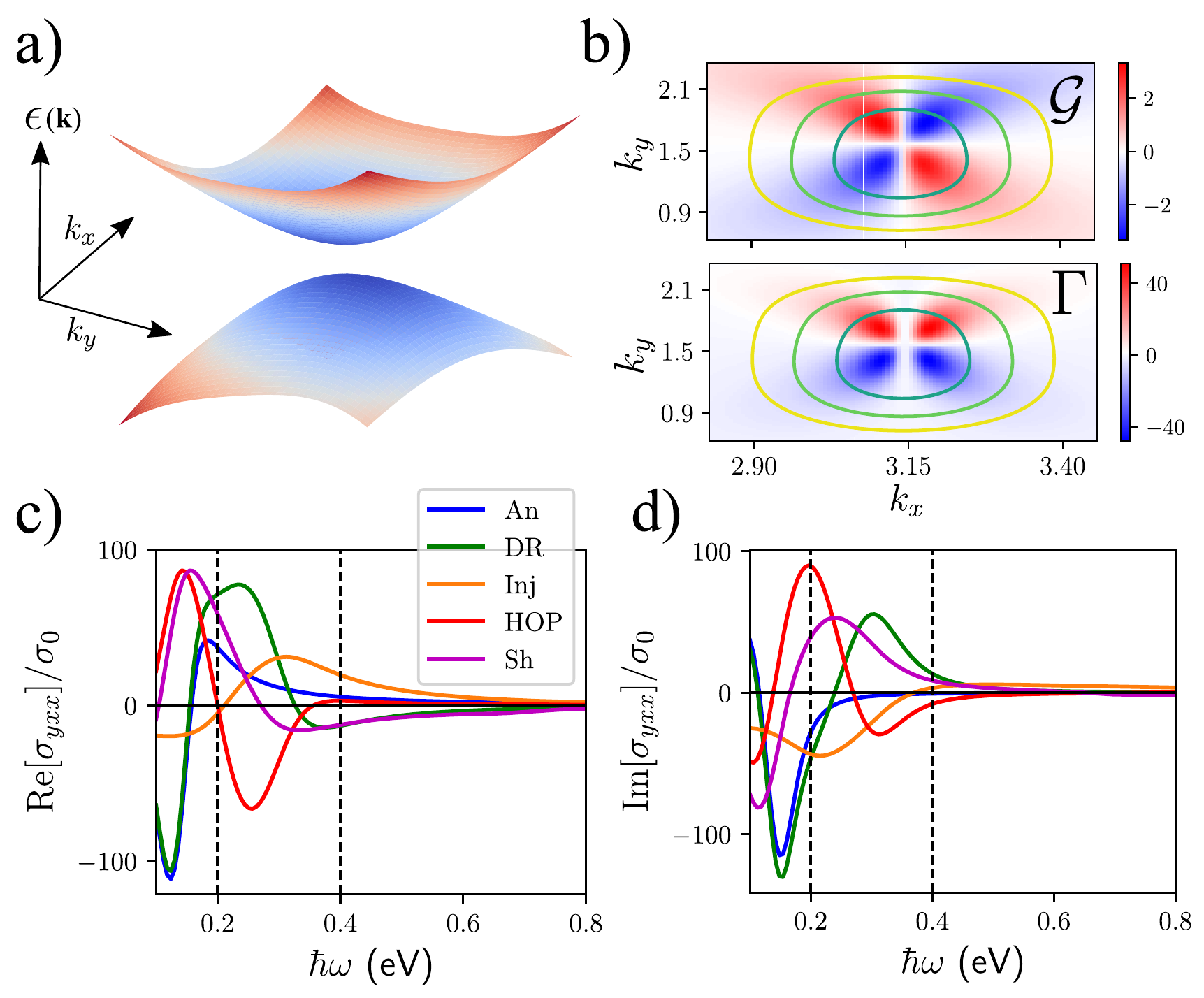}
\caption{a) Band dispersion for ${\cal P}{\cal T}$ symmetric CuMnAs [Eq.~\eqref{model_2}]. Here we set the hopping $t=0.08$ eV and $\tilde t=1$ eV. The other parameters are $\alpha_{\rm R}=0.8, \alpha_{\rm D}=0$ and ${\bm h}_{\rm AF}=(0, 0, 0.85)$ eV. b) The momentum space distribution of ${\mathcal G}^{xy}_{cv}$ (top) and the metric connection $ {\Gamma}^{yxx}_{cv}$ (bottom). Frequency dependence of the different contributions in the c) real and d) imaginary part of the SH conductivity $\sigma_{yxx}$. 
The conductivities are expressed 
in units of $\sigma_0= 10^{-5}$ ${\rm nA.m}/{\rm V}^2$, while $\mu=0.2$ eV, $\tau=1$ ps and temperature $T=12$ K.}
\label{fig_2}
\end{figure}
%

In contrast to $\mathcal T$ symmetric systems, in $\mathcal{PT}$ symmetric systems the SH generation 
i) can have a finite Drude and injection contributions, ii) have quantum geometry induced contributions are determined solely by $\mathcal G$ and $ \Gamma$, and iii) the $\cal G$ induced contributions can have diagonal components of the form $\sigma_{aaa}$. An example of a ${\cal P}{\cal T}$ symmetric material is CuMnAs, where the ${\cal P}$ and ${\cal T}$ symmetries are individually not preserved \cite{watanabe_PRX2021}. These systems generally show an anti-ferromagnetic ordering on the two distinct sub-lattice sites along with a locally broken inversion symmetry at the sub-lattice level (denoted below by A and B). This also gives rise to a sub-lattice dependent spin-	orbit coupling. The corresponding Hamiltonian is given by \cite{Jungwirth_PRL_2017, watanabe_PRX2021}
\begin{equation} \label{model_2}
\ba
\mathcal{H}({\bm k}) &\dps = 
\begin{pmatrix}
\epsilon_0 ({\bm k}) + {\bm h}_{\rm A}({\bm k}) \cdot {\bm \sigma} & V_{\rm AB}({\bm k})\\
V_{\rm AB}({\bm k}) & \epsilon_0 ({\bm k}) + {\bm h}_{\rm B}({\bm k}) \cdot {\bm \sigma}
\end{pmatrix}
\ea~.
\end{equation}
Here, we have defined $\epsilon_0({\bm k}) = -t (\cos k_x + \cos k_y)$ and $V_{\rm AB}({\bm k}) = -2 \tilde t \cos({k_x}/{2}) \cos({k_y}/{2})$.
The hopping between orbitals of the same sub-lattice is quantified by $t$, while $\tilde t$ denotes hopping between orbitals on different sub-lattices. The sub-lattice dependent spin-orbit coupling and the magnetization field is included in ${\bm h}_{\rm B}({\bm k})=-{\bm h}_{\rm A}({\bm k})$, where ${\bm h}_{\rm A}({\bm k}) = \{h_{\rm AF}^x - \alpha_{\rm R} \sin k_y + \alpha_{\rm D} \sin k_y, h_{\rm AF}^y + \alpha_{\rm R} \sin k_x + \alpha_{\rm D} \sin k_x, h_{\rm AF}^z\}$. 
The energy eigenvalues are $
\varepsilon({\bm k}) = \epsilon_0 \pm \sqrt{V_{\rm AB}^2 + h_{{\rm A}x}^2 +  h_{{\rm A}y}^2 +  h_{{\rm A}z}^2}$.  
Finite $\epsilon_0$ breaks the particle-hole symmetry and since $h_{{\rm A}x}(-k_x, -k_y) \neq h_{{\rm A}x}(k_x, k_y)$, we have $\varepsilon(-{\bm k}) \neq \varepsilon({\bm k})$. 

The dispersion in the vicinity of one of the two band edges is shown in Fig.~\ref{fig_2}(a), along with the quantum metric (top) and the metric connection (bottom) in Fig.~\ref{fig_2}(b). We find four components of the SH conductivity to be non-zero,  $\sigma_{xxy}$, $\sigma_{xyx}$, $\sigma_{yxx}$ and $\sigma_{yyy}$ with finite contributions from all the terms in Table~\ref{table_1}. The real and imaginary parts of  $\sigma_{yxx}$ are shown in Figs.~\ref{fig_2}(c)-(d). We find a finite contribution from the injection current (orange curve) which was absent in Fig.~\ref{fig_1}. Finite contributions from the DR and the HOP terms, comparable to the other contributions,  can also be clearly seen. The resonant DR peaks deviate from $\omega = \mu$ and $\omega = 2\mu$, due to the absence of particle-hole symmetry in ${\cal P}{\cal T}$ symmetric systems.

To conclude, we have unveiled two new SH phenomena, double resonant and higher order pole, which are of the same order of magnitude as previously known terms. Our work provides a detailed framework for understanding different contributions to the SH and photogalvanic currents, and their explicit dependence on the geometric properties of the electron wave-function. Our results pave the way for a full quantum geometric description of second-order non-linear optics.

\textit{Acknowledgements} - DC is supported by the Australian Research Council Centre of Excellence in Future Low-Energy Electronics Technologies (project number CE170100039). PB acknowledges the National Key Research and Development Program of China (grant No. 2017YFA0303400), China postdoctoral science foundation (grant no. 2019M650461) and NSFC grant no. U1930402 for financial support. AA and KD acknowledge IIT Kanpur - India, Science Engineering and Research Board (SERB)- India, and the Department of Science and Technology (DST) - India,  for financial support.  

{\it Note -} P.B. and K.D. contributed equally to this work.

\bibliography{DR_SHG}

\end{document}